\documentstyle[12pt]{article}
\renewcommand \baselinestretch{1.2}

\begin{document}

\def\beq{\begin{equation}}
\def\eeq{\end{equation}}
\def\bce{\begin{center}}
\def\ece{\end{center}}
\def\bea{\begin{eqnarray}}
\def\eea{\end{eqnarray}}
\def\ben{\begin{enumerate}}
\def\een{\end{enumerate}}
\def\ul{\underline}
\def\ni{\noindent}
\def\nn{\nonumber}
\def\bs{\bigskip}
\def\ms{\medskip}
\def\wt{\widetilde}
\def\wh{\widehat}
\def\tr{\mbox{Tr}\, }
\def\brr{\begin{array}}
\def\err{\end{array}}
\def\dsp{\displaystyle}


\hfill UTF 355

\hfill hep-th/9508097

\hfill July, 1995

\vspace*{1cm}

\begin{center}

{\LARGE \bf A comment on the theory of turbulence without pressure
proposed by
Polyakov}

\vspace{8mm}

\renewcommand
\baselinestretch{0.5}
\medskip

{\sc E. Elizalde}
\footnote{Permanent address:
Center for Advanced Study CEAB, CSIC, Cam\'{\i} de Santa
B\`arbara, 17300 Blanes,
Spain;
e-mail: eli@zeta.ecm.ub.es} \\
Dipartimento di Fisica, Universit\`a degli Studi di Trento,
I-38050 Povo (Trento) Italia \\
Istituto Nazionale di Fisica Nucleare, Gruppo Collegato di Trento

\vspace{15mm}

{\bf Abstract}

\end{center}

Owing to its lack of derivability, the dissipative anomaly operator
appearing in
 the theory of turbulence without pressure recently proposed by
Polyakov appears to
be quite elusive. In particular, we give arguments that seem to
lead to the conclusion that
 an anomaly in the first equation of the sequence of conservation
laws cannot be
{\it a priori} excluded.

\vspace{4mm}


\newpage

In a recent paper \cite{1} (see also \cite{2}) Polyakov has put forward
a method to treat
 turbulence with exact
field theoretical methods, in the case when the effect of pressure is
negligible.
The work has been inspired in a paper by Chekhlov and Yakhot \cite{3},
where new results
concerning Burgers' turbulence have been given. The starting point for
this
one-dimensional case is
 Burgers equation
\begin{eqnarray}
u_{t}+uu_{x}&=&\nu u_{xx} +f(xt), \nn \\
< f(x ,t) f(x' ,t') > & =&\kappa (x-x') \, \delta (t- t'),
\end{eqnarray}
where $\kappa$ is a function that defines the spatial correlation of  the
random
forces.
 For  the  generating functional
\begin{equation}
Z( \lambda_{1} x_{1} | \ldots | \lambda_{N} x_{N} ) = \left\langle
\exp{\sum{\lambda_{j}u(x_{j} t)}} \right\rangle,
\end{equation}
one obtains
\begin{equation}
\dot{Z}+ \sum{\lambda_j{\partial \over \partial \lambda_j}\left({1 \over
\lambda_j}{\partial Z\over \partial x_j}\right)} = \sum\lambda_j
\left\langle
\left[f(x_j t)  +\nu u'' \right]\exp{\sum{\lambda_{k}u(x_{k}t)}}
\right\rangle,
\nonumber\end{equation}
and further \cite{1}
\begin{eqnarray}
\dot{Z} + \sum\lambda_{j} {\partial \over \partial \lambda_{j}}
\left({1 \over
\lambda_{j}}{\partial {Z}\over \partial x_{j}}\right) = \sum{\kappa
(x_{i} - x_{j})
\lambda_{i} \lambda_{j}} Z + D,
\end{eqnarray}
where $D$ is the dissipation term
\begin{eqnarray}
D = \nu \sum\lambda_{j} \left\langle u'' (x_{j} t)
\exp{\sum{\lambda_{k} u(x_{k}t)}} \right\rangle.
\end{eqnarray}

   If the viscosity $ \nu $  were zero one would
have a  closed differential equation for $Z$. To  reach the inertial
range one
must, however, keep $\nu $ infinitesimal but non-zero. The anomaly
mechanism
mentioned above implies that infinitesimal viscosity produces a   finite
effect, whose computation is one of the main objectives in \cite{1}.
In a first stage, the  inviscid equations (5) have been considered
($\nu =0$).
Then, modulo the stirring force
and the  viscosity, one has the following sequence of conservation laws
for Eq. (1)
\begin{eqnarray}
{\partial \over \partial t}(u^{n}) +{n \over n+1} \, {\partial \over
\partial
x}(u^{n+1})\approx 0, \qquad n=1,2,3, \ldots \label{13p}
\end{eqnarray}
the sign $\approx$  meaning precisely that the
viscosity and the stirring force terms  are dropped out \cite{1}.

As discussed by Polyakov in detail, Eq. (5) can be interpreted as a
relation for
 the generating
functionals $ \left\langle u^{n_{1}}(x_{1})\ldots u^{n_{k}}(x_{k})
\right\rangle $, involving both the stirring force and the viscosity. The
 latter presents a problem. The
rule  is that in any equation involving  space points separated by
a distance larger than $a$, the viscosity can be put equal to zero.

And here comes the specific situation we want to deal with.  In
principle, it
 seems
legitimated to use the inviscid limit for the first equation,
$ n=1$, of (\ref{13p}),
 because in this case one can make use of
the  steady state condition
\beq
{d \over dt}{\left\langle u(x_{1}) \ldots u(x_{N}) \right\rangle}=0,
\qquad
 \mid x_{i}-x_{j}\mid \gg a.
\eeq
The problems seems to start with the case $n=2$, because
then one has to
take a time derivative of the product of two (or more) $ u$'s  at the
same point.
This problem has been solved, in the case $n=2$, by making the
replacement
\beq
u^{2}(x)\Longrightarrow u(x+{y \over 2})u(x-{y \over 2}), \qquad \mid
x_{i}-x_{j}\mid \gg y\gg a
\eeq
 and by leting $ y\rightarrow 0$ only after the viscosity is taken to
zero.
Using then the inviscid equations for $n=1$ one can write
\beq
-{d \over dt}{\left[u(x_1)u(x_2)\right]}\approx {1 \over 2}
\left\{ {\partial
\over \partial x_{1}} \left[u^{2}(x_{1})u(x_{2})\right]+ {\partial
\over \partial x_{2}} \left[u^{2}(x_{2})u(x_{1})\right]\right\}, \quad
 x_{1,2}=x\pm {y \over 2}.
\eeq
By employing simple algebraic identities, and the following expression
\beq
 {\partial \over \partial y}{\left[u^{3}(x_1)-u^{3}(x_2)\right]}={1
\over 2}{\partial \over \partial x}{\left[u^{3}(x_1)+u^{3}(x_2
)\right]} \ \longrightarrow  \ {\partial \over \partial x} u^{3}(x),
\label{ex1}
\eeq
one gets
\beq
-{d \over dt}{\left[u(x_1)u(x_2)\right]}
 \approx{2
\over3}{\partial \over \partial x}{u^{3}(x)}+ a_{0}(x),
\eeq
where $ a_{0}(x)$ is a dissipative  anomaly operator, given by \cite{1}
\beq
a_{0}(x)=\lim_{y\rightarrow 0}{{1 \over 3}{\partial \over \partial y}
{\left[u(x_1
)-u(x_2)\right]^{3}}}.
\eeq
It is here crucial to observe that the anomaly  would be zero if $ u(x)$
were
 differentiable.  However, as remarked in \cite{1}, the steady state
 condition clearly prevents this from being true. Indeed, one of the
consequences
 of  Eq. (5) is that in the steady-state situation one has
\begin{eqnarray}
{d \over dt}{\left\langle u^{2} \right\rangle}=\kappa (0) -\left\langle
a_{0}
\right\rangle =0,
\end{eqnarray}
and  the celebrated Kolmogorov relation holds
\begin{eqnarray}
\left\langle \left[u(x_{1})- u(x_{2})\right]^{3} \right\rangle \propto
\kappa
(0)(x_{1}-x_{2}).
\end{eqnarray}
The value of the anomaly defines the limiting contribution of the viscous term
in the steady state
\beq \lim_{\nu\rightarrow 0} \, {\nu u(x)u''(x)} = -a_{0}(x).
\eeq

Notice, again, that the fact that the anomaly $a_0(x)$ is non vanishing
(together with
 its important
consequences, as the Kolmogorov relation) depends solely on the
non-differentiability of the
function $u(x)$. Simple considerations ---the first of which could
be pure symmetry---
can lead us easily to the conclusion that an anomaly of the same
type can be also present in the
first of the equations. In fact, its absence has not been proven
in \cite{1}, but just the
compatibility of the general argument with the fact that it can be zero
(the whole argument
 has been
qualified by Polyakov himself as a {\it consistent conjecture}
\cite{1}).

Crude symmetry considerations would yield
\beq
-{d \over dt} \, u(x)
 \approx{1 \over2}{\partial \over \partial x}{u^2(x)}+ \lim_{y\rightarrow 0} {1
 \over 2} {\partial \over \partial y} \left[u(x_1
)-u(x_2)\right]^2, \label{n1}
\eeq
where, again the non-differentiability of the function $u(x)$ permits the
anomaly term
 (the second one on the rhs) to be non-zero.
It is easy to see that if one exactly parallels the derivation leading
to the presence of $a_0$, this term does not appear \cite{5}. However,
the main argument that leads to the possibility of an anomaly in the
first equation is actually of the same kind, namely that
 owing to the non-differentiability of $u(x)$,
pretended identities as the last step in Eq. (\ref{ex1}) might acquire
an additional contribution. One should observe that by using the
Burgers equation any discontinuity in the time derivative of $u$
 and its powers can be traced back to a corresponding discontinuity in
the spatial derivative.
But then,
point-splitting the $x-$derivative of $u^2(x)$ (what is {\it not} trivial
matter,
 given the non-differentiability of $u(x)$), in the form
\beq
{1 \over 2} {\partial \over \partial x} \, u^2(x)
 \longrightarrow {1 \over 2}{\partial \over \partial
x}\left[u(x_1)u(x_2)\right], \eeq
and using the same kind of manipulations as in \cite{1}, we are led to the
possible presence of an anomaly which is {\it not} exactly given by
the immediate guess (\ref{n1}) but by a very similar expression:
\beq
{1 \over 2}{\partial \over \partial x}\left[u(x_1)u(x_2)\right]
\longrightarrow {1 \over 4} \left[
{\partial \over \partial x} \, u^2(x_1) +
{\partial \over \partial x} \, u^2(x_2) \right] -
\left[ u(x_1)-u(x_2)\right] {\partial \over \partial y}
\left[ u(x_1)+u(x_2)\right].
\eeq
Using the same labeling of Polyakov, we will call it
$a_{-1}(x)$,
\beq
a_{-1}(x)=\lim_{y\rightarrow 0} \
\left[ u(x_1)-u(x_2)\right] {\partial \over \partial y}
\left[ u(x_1)+u(x_2)\right].
\eeq
It is seen to be non-vanishing in general. This is realized by direct
calculation of the
derivative as a quotient of differences and by
considering several possible ways of taking the
two limits involved, namely the one of the derivative
itself and the limit $y \rightarrow 0$.
Notice that this non-vanishing is, naively, even more strong than in the
cases $n=2$
 and further, because
differentiability would here yield an infintesimum of
first order only, while in the case
 $n=k$ it would be of the corresponding order $k$.

It seems difficult to give an immediate physical meaning to this
possible anomaly.
In the case of the anomalies $a_0, a_1, \ldots$, Polyakov has produced a
beautiful interpretation, as limiting contributions of the viscous term
in the steady-state situation $\frac{d}{dt} \langle u^n \rangle =0$:
\beq
a_0(x) = - \lim_{\nu \rightarrow 0} \nu u(x) u''(x),
\eeq
and similarly for the others. As observed by Falkovich \cite{4}, due to
the fact that $\int u dx$ is an integral of motion of the complete
Burgers equation, we do not get a steady-state condition for the case
$n=1$. This means in some way that the corresponding equality for
$a_{-1}$,
\beq
a_{-1}(x) = - \lim_{\nu \rightarrow 0} \nu u''(x),
\eeq
should be a universal one but, on the other hand, $\langle a_{-1} (x)
\rangle =0$. This trivial situation  obviously ceases to be so if one
allows the viscosity $\nu$ not to be strictly constant.

As is plain, such possibility would bear with it
important consequences, as a
corresponding Kolmogorov relation (with power 2 instead of power 3), and
 it would even  modify the
Kolmogorov relation itself
and all the subsequent anomalies (starting from $a_0(x)$), since it would
 contribute a term in the
derivation of the relations for $u^2$ and all the subsequent $u^n$.
To finish, let us stress again that the existence of $a_{-1}$ and of
all these additional
terms is, in principle, just a possibility, brought about
in a natural way by the same mathematical
argument that leads to the introduction of the anomalies
$a_n$, $n \geq 0$, consistently.

 \vspace{5mm}


\noindent{\large \bf Acknowledgments}

It is a pleasure to thank the members of the Department of Theoretical
Physics of the
 University of Trento for
warm hospitality, specially Sergio Zerbini, Luciano Vanzo, Guido Cognola,
Ruggero Ferrari
 and Marco Toller. Comments by Grisha Falkovich and Rahimi Tabar are
also gratefully acknowledged.
This investigation has been supported by DGICYT (Spain), by CIRIT
 (Generalitat de Catalunya) and by the INFN (Italy).

\end{document}